\documentclass[conference]{IEEEtran}
\usepackage[pdftex]{graphicx}
\usepackage[usenames,dvipsnames,table,xcdraw]{xcolor}
\usepackage{array}
\usepackage{multirow}
\usepackage[normalem]{ulem}
\useunder{\uline}{\ul}{}
\usepackage{booktabs}
\usepackage{supertabular,booktabs}
\usepackage{longtable}
\usepackage{lscape}
\usepackage{cite}
\usepackage{xurl}
\usepackage{dirtytalk}
\usepackage{graphicx}
\usepackage{comment}
\usepackage{makecell}
\usepackage{svg}
\usepackage{amsmath}

\usepackage[breaklinks]{hyperref} 

\usepackage[caption=false]{subfig}
\graphicspath{{Images/}}
\usepackage[normalem]{ulem}
\useunder{\uline}{\ul}{}

\usepackage{tabularx}
\usepackage{balance}
\usepackage{nameref}
\newcommand{\MyBox}[1]{\vspace{3mm}\noindent\framebox[\columnwidth][c]{\parbox[b]{0.95\columnwidth}{ #1 }}\vspace{3mm}}

\begin{document}

\title{Software Fairness Testing in Practice}

\author{
\IEEEauthorblockN{Ronnie de Souza Santos}
\IEEEauthorblockA{University of Calgary\\
Calgary, AB, Canada \\
ronnie.desouzasantos@ucalgary.ca}
\and

\IEEEauthorblockN{Matheus de Morais Leça}
\IEEEauthorblockA{University of Calgary\\
Calgary, AB, Canada \\
matheus.demoraisleca@ucalgary.ca}

\and

\IEEEauthorblockN{Reydne Santos}
\IEEEauthorblockA{UFPE\\
Recife, PE, Brazil \\
rbs8@cin.ufpe.br}
\and

\IEEEauthorblockN{Cleyton Magalhaes}
\IEEEauthorblockA{UFRPE\\
Recife, PE, Brazil \\
cleyton.vanut.ufrpe.br}
}


\IEEEtitleabstractindextext{%
\begin{abstract}
Software testing ensures that a system functions correctly, meets specified requirements, and maintains high quality. As artificial intelligence and machine learning technologies become integral to software systems, testing has evolved to address their unique complexities. A critical advancement in this space is fairness testing, which identifies and mitigates biases in AI applications to promote ethical and equitable outcomes. Despite extensive academic research on fairness testing, including test input generation, test oracle identification, and component testing, practical adoption remains limited. Industry practitioners often lack clear guidelines and effective tools to integrate fairness testing into real-world AI development. This study investigates how software professionals test AI-powered systems for fairness through interviews with 22 practitioners working on AI and ML projects. Our findings highlight a significant gap between theoretical fairness concepts and industry practice. While fairness definitions continue to evolve, they remain difficult for practitioners to interpret and apply. The absence of industry-aligned fairness testing tools further complicates adoption, necessitating research into practical, accessible solutions. Key challenges include data quality and diversity, time constraints, defining effective metrics, and ensuring model interoperability. These insights emphasize the need to bridge academic advancements with actionable strategies and tools, enabling practitioners to systematically address fairness in AI systems.
\end{abstract}

}

\maketitle

\IEEEdisplaynontitleabstractindextext

\IEEEpeerreviewmaketitle

\section{Introduction}
\label{sec:introduction}

Software testing has changed significantly over the past decades, always evolving to support new challenges and technological advancements \cite{gillenson2018literature, garousi2016systematic}. Initially, testing was indistinguishable from debugging, relying on ad hoc methods to fix defects. By the 1950s-1970s, testing became a distinct practice focused on verifying correctness while debugging addressed failures. The 1970s-1980s saw a shift toward error detection, followed by the 1980s-1990s, when formal standards emphasized early defect prevention. In the post-1990s era, methodologies like Test-Driven Development embedded testing into the software lifecycle \cite{gillenson2018literature, gelperin1988growth}. Today, with the rise of machine learning and AI, testing has further evolved to address the non-deterministic nature of AI systems \cite{gao2019ai, tao2019testing}. 

In the current landscape of software engineering, fairness testing has emerged as a key component of quality assurance, addressing the ethical challenges posed by AI-driven systems. These systems, capable of performing complex tasks such as recommendation, object detection, classification, and prediction, require advanced validation techniques to ensure their reliability and fairness \cite{brun2018software, verma2018fairness}. Unlike traditional software, AI models continuously adapt to data, making them susceptible to biases introduced through training data, algorithmic design, or societal inequalities \cite{desoftware}. Fairness testing, then, aims to systematically evaluate the ethical implications and potential biases embedded within AI systems, ensuring that decision-making processes lead to equitable outcomes across diverse demographic groups. By identifying and mitigating discriminatory risks based on factors such as race, gender, or socioeconomic status, fairness testing plays a key role in ensuring ethical AI development and promoting fairness in automated decision-making \cite{galhotra2017fairness, aggarwal2019black}.

While extensive research has been conducted on fairness testing, focusing on areas such as test input generation, test oracle identification, and the testing of various components, including data testing, machine learning program testing, and model testing \cite{chen2024fairness}, much of this work remains confined to academic settings, offering limited practical guidance for industry professionals. This gap leaves practitioners struggling to effectively implement fairness testing and manage fairness-related bugs in real-world AI systems \cite{galhotra2017fairness}. For example, there is a growing demand for fairness testing tools that align more closely with industry needs, particularly in terms of usability, comprehensive coverage, and configurability, ensuring they can be seamlessly integrated into existing development workflows \cite{nguyen2024literature}.

Therefore, considering that fairness testing is a relatively new aspect of software development practice, and that theories and academic findings often take time to be adopted by industry, this paper investigates how software professionals are testing AI-powered systems and what strategies they use to verify biases in the software. To guide our research, we formulated the following question:

\smallskip \smallskip
{\narrower \noindent \textit{\textbf{RQ.} How are software professionals currently testing AI-powered systems for fairness and bias?} \par}
\smallskip \smallskip

Answering this question has direct practical implications, helping software professionals grasp the current state of fairness testing in AI systems while enabling academic researchers to identify gaps between theory and industry practice. By bridging this divide, our study highlights areas where further research and development are needed to align academic insights with real-world implementation. This alignment can drive the creation of more effective tools, methods, and guidelines to tackle bias in software systems. In summary, our study provides four key contributions:

\begin{enumerate}
    \item It identifies the primary focus areas of fairness testing in the industry, highlighting diversity and inclusivity, model consistency, and data representation and balancing as key testing targets.
    \item It shows that fairness testing in practice currently relies heavily on ad-hoc and iterative approaches rather than standardized methods.
    \item It highlights key challenges in fairness testing, including poor data quality and diversity, time constraints, lack of specialized tools and metrics, fairness knowledge gaps within teams, and black-box behavior issues.
    \item It reports the need for scalable, automated tools, clearer industry guidelines, and stronger cross-functional collaboration between AI developers and testing professionals. 
\end{enumerate}

From this introduction, this study is organized as follows. In Section 2, we present a literature review on software fairness and fairness testing. Section 3 describes the method conducted in this study. In Section 4, we present our findings, which are discussed in Section 5, along with the implications and limitations of this study. Finally, Section 6 summarizes our contributions and final considerations.

\section{Background} \label{sec:background}
This section is focused on exploring two concepts central to this study: software fairness and fairness testing.

\subsection{Software Fairness}

Although there is no single, universally accepted definition of fairness in the software context, the term generally refers to a set of properties designed to prevent discriminatory behaviors or biases, particularly regarding sensitive attributes such as gender, race, and age \cite{kwiatkowska1989survey, soremekun2022software, verma2018fairness}. Ensuring fairness is essential for all types of software systems, especially AI-driven applications used in critical domains like criminal justice, healthcare, and finances \cite{brun2018software}. Without proper fairness considerations, these systems can perpetuate or even amplify existing social inequalities, leading to biased and discriminatory outcomes that disproportionately impact certain individuals or groups \cite{soremekun2022software}.

The absence of fairness in software systems can have far-reaching consequences, reinforcing and perpetuating existing social inequalities \cite{soremekun2022software}. Unfair systems often amplify biases present in training data, leading to discriminatory outcomes that disproportionately impact marginalized individuals or groups \cite{soremekun2022software, brun2018software, mehrabi2021survey}. Beyond social harm, fairness issues can also damage organizational reputation and reduce public trust in automated systems. The deployment of biased software can lead to negative perceptions of the organizations that use it, reducing confidence in AI-driven decision-making. Additionally, fairness violations carry legal and ethical risks, particularly in sensitive domains such as criminal justice and healthcare, where biased decisions can have severe real-world consequences \cite{brun2018software}.

Ensuring fairness in software presents significant challenges due to biased data, software complexity, and the lack of dedicated tools and methods. AI systems inherit biases from their training data, reinforcing inequalities if those biases go unaddressed \cite{brun2018software}. The increasing complexity of software, especially machine learning-based systems, makes bias identification and mitigation difficult, as intricate component interactions and lack of transparency in algorithms can lead to unexpected biases \cite{brun2018software, mehrabi2021survey}. Additionally, few tools and methods are currently available to assess and ensure fairness, with research in fairness testing, debugging, and formal verification still evolving \cite{brun2018software}. Fairness in software extends beyond technical concerns—it is also an ethical and social issue, requiring collaboration with ethicists, social scientists, and affected communities to develop AI systems that promote fairness and inclusivity \cite{brun2018software, desoftware}.

\subsection{Fairness Testing}
Despite efforts to design discrimination-free algorithms, bias often persists due to implementation errors, emergent system behaviors, and automated learning from data \cite{galhotra2017fairness}. This makes fairness testing essential for identifying and mitigating system bias \cite{falk2008testing}. Fairness Testing is the process of verifying and validating the software to reveal fairness bugs, which are flaws that cause discrepancies between actual behavior and fairness conditions \cite{chen2024fairness}. In other words, it ensures that software operates impartially, without discriminating against individuals or groups based on sensitive attributes such as gender, race, or age \cite{chen2024fairness}.

The growing reliance on automated systems has heightened interest in fairness testing, as recent studies have highlighted failures in AI-driven decision-making \cite{letzter2016amazon, angwin2022machine}. The literature outlines key steps in the process of fairness testing \cite{chen2024fairness}: 
\begin{enumerate}
    \item Defining fairness conditions, where engineers specify fairness expectations.
    \item Identifying and creating fairness oracles, which determine if software behavior aligns with these conditions.
    \item Obtaining test inputs by sampling or generating data.
    \item Executing the test inputs to detect fairness violations.
    \item Assessing test adequacy, evaluating whether fairness bugs are effectively identified.
    \item Generating bug reports to aid in reproduction and resolution.
    \item Implementing software repairs to correct fairness violations.
    \item Monitoring the system to ensure continued compliance with fairness conditions.
\end{enumerate}

Fairness testing can be applied to various software components, including training data, machine learning models, algorithms, frameworks, and non-ML components \cite{chen2024fairness, udeshi2018automated}. However, although fairness testing techniques exist, several challenges remain. Defining and measuring fairness is complex, as different contexts require different fairness definitions, making it difficult to establish universal evaluation standards \cite{galhotra2017fairness}. Even when fairness conditions are defined, quantifying bias and setting violation thresholds can be highly context-dependent \cite{galhotra2017fairness, udeshi2018automated}. Additionally, generating meaningful test cases poses difficulties, as modifying sensitive attributes can result in unrealistic tests and false positives \cite{chen2024fairness}. Finally, selecting the appropriate fairness oracle is challenging, as different fairness metrics may be more relevant depending on the application domain \cite{chen2024fairness}.

\section{Method} \label{sec:method}
In this study, we conducted a case study to investigate how software professionals deal with fairness testing in their daily work in software projects that involve AI development. A case study is a research method that examines contemporary phenomena in real-world contexts using multiple sources of evidence. In software engineering, case studies are valuable for studying complex, context-dependent processes such as development and practices \cite{runeson2009guidelines}. By observing real-world practices, case studies provide insights into how software engineers work under different conditions, addressing social, technical, and organizational factors \cite{runeson2009guidelines, wohlin2021case}. In this study, we followed well-established guidelines to conduct case studies in software engineering \cite{ralph2020empirical, runeson2009guidelines, easterbrook2008selecting}. 

\subsection{The Case} \label{sec:context}

The case under study is a large South American software company established in 1996 that provides software solutions across various sectors, including finance, telecommunications, government, manufacturing, services, and utilities. Of its 1,200 employees, over 70\% are directly engaged in software development, spread across over 70 teams. These teams are composed of professionals with diverse technical expertise (e.g., programmers, quality assurance engineers, designers) and personal backgrounds (e.g., gender and ethnic diversity). They are familiar with several development methodologies, such as Scrum, Kanban, and Waterfall, and deliver solutions to clients across North America, Latin America, Europe, and Asia.

Using semi-structured interviews as our main data collection approach, we interviewed a group of software professionals directly involved in the development of AI and machine learning-based software solutions, including designers, software engineers, data scientists, and testers, from four different projects: Project A developed a real-time application that translates sign language into Portuguese text using deep learning neural networks \cite{mari2020libras}. Project B involved creating digital twins and predictive models for the petroleum and energy sector, Project C utilized large language models (LLMs) in education, and Project D employed computer vision for facial recognition \cite{shaikh2025lip}). While specific details about Projects B and C are restricted due to confidentiality agreements, all projects, regardless of their individual or industrial focus, share a reliance on AI technologies. This reliance highlights the need to address bias issues and ensure fairness, or bias testing is implemented to avoid discriminatory or biased outcomes and ensure trustworthy AI systems.

\subsection{Participants} \label{sec:participants}
To select participants for the study, we initially used convenience sampling \cite{baltes2022sampling}, selecting individuals based on availability and any contractual constraints. We sent an open invitation to all professionals working on the selected projects, asking those interested to indicate their preferred dates and times for interviews. To expand the participant pool, we employed snowball sampling \cite{baltes2022sampling}, inviting interviewees to recommend colleagues who might also be interested and able to contribute.

To further diversify perspectives and address key research questions, we then applied theoretical sampling \cite{charmaz2014constructing}. This involved targeting individuals with specific expertise, experience levels, or demographic backgrounds relevant to emerging themes identified in early analyses, such as the impact of diverse backgrounds on identifying and mitigating bias in machine learning models. This multi-stage sampling strategy allowed us to gather a diverse and relevant set of participants, ensuring a broad range of insights into the studied phenomena.

Initially, we focused our interviews on software testers, as they are typically responsible for testing software and ensuring its quality. However, as data collection progressed, we observed that testers were not the only professionals involved in ensuring fairness through validation processes. For instance, we gathered evidence that data scientists were conducting tests to evaluate models and datasets, programmers were engaged in unit testing activities, and designers were testing user interfaces and interactions for potential biases. Recognizing the diverse roles contributing to fairness testing, we expanded our participant pool to include various professionals involved in the projects (through theoretical sampling \cite{charmaz2014constructing}). This allowed us to explore different perspectives on fairness testing, capturing a more comprehensive understanding of the phenomena.

\subsection{Data Collection} \label{sec:collection}
As part of our data collection process, we used a semi-structured interview guide (Table I) that evolved across three interview rounds. In the initial stage, conversations focused on general concepts of bias in AI development. As new themes emerged, subsequent rounds incorporated more targeted questions about fairness testing, such as challenges related to integrating fairness considerations into development cycles, methods used to detect bias, the roles responsible for fairness validation, and how practitioners address fairness concerns throughout the software lifecycle. This iterative refinement enabled us to capture a deeper understanding of how fairness testing is operationalized in practice and the obstacles professionals encounter in real-world projects.

Regarding the interviews, in the first two rounds, we interviewed seven participants each, with an additional eight participants joining in the third round, bringing the total to 22 professionals. We stopped conducting new interviews once we reached data saturation, where no new themes or information were emerging. The interviews about testing, held between June 1 and July 5, 2024, lasted between 15 and 25 minutes, resulting in over 200 pages of transcribed data. Each transcript was anonymized before analysis and segmented by question to align with our coding schema. To enhance organization, transcripts were indexed with participant codes and interview dates. These structured transcripts were then imported into our analysis environment for GPT-assisted and manual coding.

Throughout the interviews, we consistently took observational notes to complement the transcribed data. These notes included reflections on participants’ tone, pauses, gestures, and expressions of certainty, confusion, or frustration. Such contextual observations were used to enrich our interpretation of the transcripts and to flag moments of ambiguity or emphasis that could inform coding decisions. While these notes were not formally coded, they served as analytic memos during the analysis process.

\vspace{-0.1in}
\begin{table}[h!]
\label{interview}
\caption{Interview Guide}
\centering
\begin{tabular}{p{7.5cm}}
\hline
\textbf{\newline GENERAL QUESTIONS} \\

1. Can you describe your day-to-day work in the development of an AI system?

\textbf{\newline IDENTIFYING BIAS IN AI SYSTEMS} \\

2. How do you identify bias in the system while working on this project? \\

3. What impact does bias have on your system?

\textbf{\newline MANAGING BIAS IN AI SYSTEMS} \\

4. If you identify bias in your project, what steps do you take to manage or mitigate it? \\

5. Who is responsible for testing bias in your project? \\
\textbf{\newline FAIRNESS/BIAS TESTING PROCESS} \\

6. How is fairness or bias testing conducted in your project? Can you describe the step-by-step process used? \\

\hline
\end{tabular}
\end{table}

\vspace{-0.1in}

\subsection{Triangulation}
Triangulation was conducted through two complementary strategies. First, cross-case triangulation involved comparing findings across four AI projects (A–D), each using different technologies and addressing unique fairness concerns. Project A focused on sign language translation and addressed bias related to accessibility and representation. Project B developed digital twins for the energy sector, considering bias in predictive modeling. Project C applied large language models in education, emphasizing biases in textual data. Project D used computer vision for facial recognition, dealing with demographic bias and ethical implications. This comparative analysis enabled us to identify shared practices and challenges despite domain-specific variations.

Second, we adopted data triangulation by combining interview narratives with supporting artifacts, including test plans and internal documentation related to fairness and bias evaluations. These materials were provided by participants to enhance the depth of our analysis and allowed us to validate and cross-reference interview statements with concrete examples of fairness testing in practice.

\subsection{Data Analysis} \label{sec:analysis}
We conducted a thematic analysis \cite{cruzes2011recommended} to explore the narratives gathered from participants' interviews. Thematic analysis is a widely used qualitative approach that enables researchers to systematically identify, organize, and interpret patterns within data. This method provides a flexible yet structured framework for uncovering meaningful insights from participants' experiences, capturing both explicit information and underlying themes that reveal deeper perspectives relevant to the study’s objectives.

\begin{figure}
    \centering
    \includegraphics[width=\linewidth]{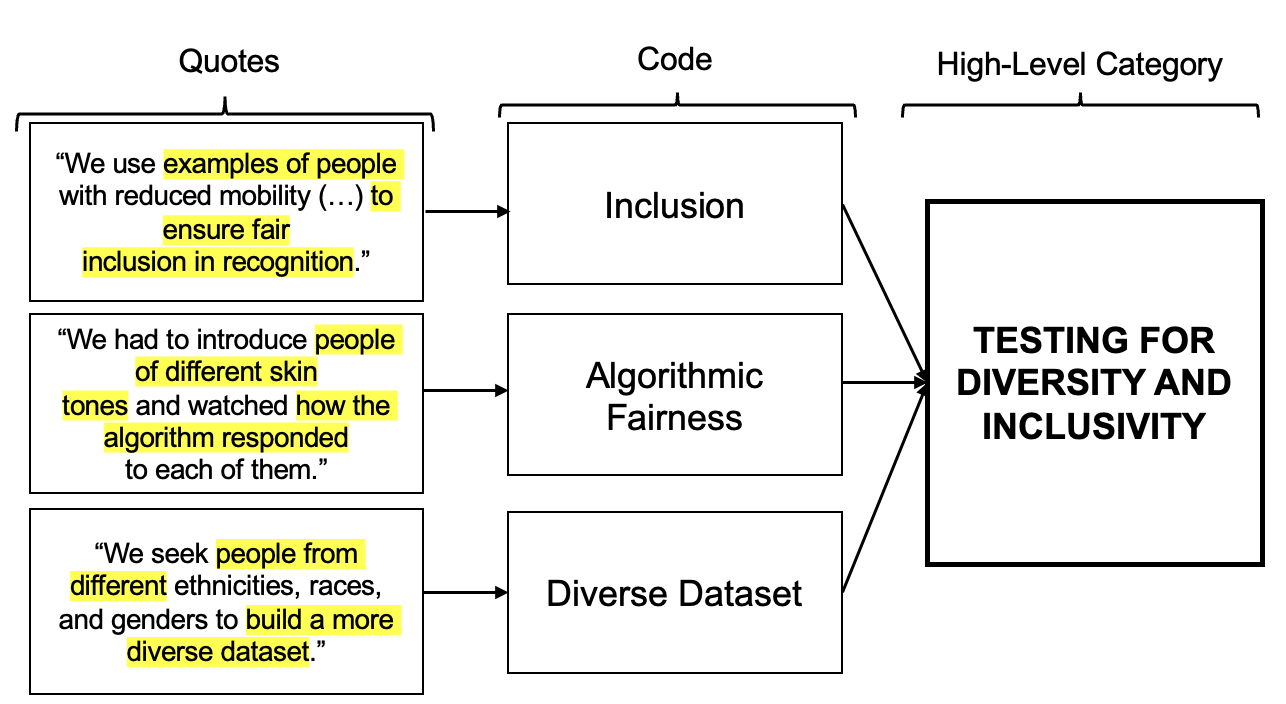}
    \caption{Thematic Analysis: Coding Process}
    \label{fig:holistic}
\end{figure}

We followed a three-phase thematic analysis approach as demonstrated in Figure \ref{fig:holistic}: (1) Open coding, where GPT-4 Omni was prompted to extract fairness testing references across transcripts; (2) Axial coding, where we manually grouped codes into broader categories based on semantic similarity and development context; and (3) Selective coding, where we synthesized these categories into broad themes.

We began with open coding, examining transcripts to identify AI testing practices and generate initial codes that described different aspects of the data. To improve efficiency in this phase, we applied GPT-4 Omni, following guidelines for using LLMs in qualitative analysis \cite{xiao2023supporting, lecca2024applications}. Leveraging an LLM aimed to significantly accelerate the coding process, acknowledging that fully manual thematic analysis is typically time-consuming \cite{taylor2018can}. We used prompt engineering (Table II) to systematically extract pertinent quotations from interview transcripts. This process involved refining the prompt and manually verifying the results to ensure accurate identification and categorization of mentions related to fairness testing \cite{Heston2023, white2023prompt, sahoo2024systematic}. The prompt presented in Table II represents the finalized version, optimized by applying multiple prompt engineering techniques \cite{lecca2024applications}. Manual review was performed to verify GPT-generated codes. After achieving consistency on our dataset, we conducted full-scale extraction. For validation, 20\% of extracted quotes were randomly selected and manually cross-checked against original transcripts to confirm accuracy. 

After completing open coding, we proceeded to axial coding, where we manually refined and grouped related codes into broader categories representing recurring patterns in the data. This phase focused on how fairness testing was performed, its integration into development workflows, the techniques used, and the challenges encountered. To enhance reliability, two authors were paired throughout the axial coding process to independently review and organize the codes. This pairing fostered discussions that helped resolve ambiguities, align interpretations, and ensure consistency across coding decisions.

Finally, in selective coding, we identified central themes and high-level categories describing fairness testing in practice, including the aspects of AI systems tested for fairness, the methods employed, and the challenges practitioners faced. These final themes provided a structured synthesis of fairness testing approaches and their real-world applications, offering valuable insights into how practitioners integrate fairness considerations into software development.

\begin{table}[h!]
\label{tab:prompt}
\caption{Data Extraction Prompt}
\centering
\scriptsize
\begin{tabular}{p{7.5cm}}
\hline
\textbf{\newline PROMPT} \\

\textbf{Instruction:} You are an experienced researcher specializing in fairness testing practices. Your task is to thoroughly and exhaustively analyze transcripts of interviews with software professionals to extract detailed information related to their testing methods, with a particular focus on AI system bias identification. Your extraction must be exhaustive, capturing all relevant details under the following aspects: \\

\textbf{i) What they or their team test:} Identify everything the professionals or their team consider important to test. \\

\textbf{ii) How they or their team test:} Describe the testing strategies, tools, or methods used, capturing all variations and nuances. \\

\textbf{iii) Their or their team’s experiences with testing:} Highlight all relevant anecdotes, challenges, successes, or reflections. \\

\textbf{iv) What they or their team avoid testing:} Identify all areas deliberately left untested or neglected, providing comprehensive details. \\

\textbf{v) Challenges they or their team face with testing:} Extract every mention of difficulties encountered, whether technical, organizational, or resource-related. \\

\textbf{\newline Output Example}

\begin{enumerate}
    \item Participant Code: P01
    \item Aspect of Testing Discussed: What they test
    \item Exact Quote: "We typically test the model's ability to generalize across different demographic groups."
\end{enumerate}

\textbf{\newline Additional Details} \\

The research question is: \textbf{How are software professionals testing AI systems for bias identification?} Extract as many relevant quotes as possible. The analysis should be exhaustive, covering every possible mention about the research. \\

\hline
\end{tabular}
\end{table}

\subsection{Ethics} 
Our research adhered to the ethical guidelines of the first author's university. Participants were informed of the research's purpose, the voluntary nature of their participation, and their right to withdraw at any time. Informed consent was obtained before each interview, and all data presented in this paper was anonymized to protect participant confidentiality. To further ensure anonymity, we removed any mention of company names, project names, or individuals referenced by interviewees before applying the prompt for data extraction.

\section{Findings} \label{sec:findings}
As described in Section~\ref{sec:participants}, we used a combination of convenience sampling, snowball sampling, and theoretical sampling to select participants for this study. This approach enabled us to recruit 22 professionals actively involved in the development of AI-powered systems, spanning a variety of technologies, including deep learning neural networks, prediction models, large language models, and computer vision applications for facial recognition. The participants represented a range of roles, such as data scientists, software developers, QA/testers, designers, and project managers, giving us a broad perspective on the processes behind AI system development.

In line with the study’s focus on fairness testing, we prioritized recruiting participants from diverse backgrounds, including those from equity-deserving groups, as research has consistently emphasized the role of diversity in bias recognition and enhancing fairness in software development ~\cite{adams2020diversity, aggarwal2019black}, making the diversity within our sample essential for understanding how team composition influences the testing process. Ensuring a varied group of participants allowed us to gain deeper insights into how diverse teams approach fairness testing and address potential biases in AI systems.

The final demographic breakdown of our sample reflects this diversity. Approximately 32\% of participants were non-male professionals, including one non-binary individual. Additionally, 8\% of participants had disabilities, with two individuals reporting a disability. The sample also included 16\% non-white professionals, with representation from mixed-race and Black individuals. 20\% identified as LGBTQIA+, comprising gay, lesbian, and queer participants. Furthermore, 12\% were neurodivergent, including individuals with ADHD. In terms of professional expertise, 36\% held advanced degrees (Master’s or Ph.D.), and 40\% had more than five years of experience in software development. This diverse and highly skilled group provided valuable perspectives on how diversity influences AI development and helps address bias. Further details are provided in Table \ref{tab:Demographics}.

\begin{table}
\centering
\caption{Demographics}
\renewcommand{\arraystretch}{1}
\label{tab:Demographics}
\begin{tabular}{llr}
\toprule
\multirow{2}{*}{ \textbf{Gender} } 
& Men & 14 individuals\\
& Women & 7 individuals\\ 
& Non-binary & 1 individual \\ \midrule 

\multirow{6}{*}{ \textbf{Role} } 
& Data Scientists & 9 individuals\\
& Designers & 4 individuals\\
& Programmers & 3 individuals\\
& Testers & 4 individuals\\
& Researchers & 1 individual\\
& Managers & 1 individual \\ \midrule 

\multirow{4}{*}{ \textbf{Education} } 
& Bachelor's Degree & 9 individuals\\ 
& Postbaccalaureate & 5 individuals\\ 
& Master's Degree & 7 individuals\\ 
& PhD Degree & 1 individual\\ \midrule 

\multirow{4}{*}{ \textbf{Experience} }  
& 1-3 years & 3 individuals\\ 
& 3-5 years & 9 individuals\\ 
& 5-10 years & 6 individuals\\ 
& More than 10 years & 4 individuals\\ \midrule 

\multirow{3}{*}{ \textbf{Ethnicity} } 
& White & 18 individuals\\
& Mixed-race & 3 individuals\\
& Black & 1 individual \\ \midrule 

\multirow{2}{*}{ \textbf{Disability} } 
& Without & 20 individuals\\
& With & 2 individuals \\ \midrule 

\multirow{2}{*}{ \textbf{LGBTQIA+} } 
& No & 17 individuals\\
& Yes & 5 individuals \\ \midrule 

\multirow{2}{*}{ \textbf{Neurodivergent} } 
& No & 19 individuals\\
& Yes & 3 individuals\\
\bottomrule
\end{tabular}
\end{table}

\subsection{Fairness Testing in Practice: What is Being Tested}
\begin{table}[h!]
\caption{Targets for Fairness Testing}
\centering
\label{tab:fairness_testing_targets}
\begin{tabularx}{\linewidth}{p{2cm} X}
\toprule
\textbf{Target} & \textbf{Evidence Examples} \\ \hline

\textbf{\textit{Testing for \newline Diversity and \newline Inclusivity}} & \textit{``We use examples of people with reduced mobility and other types of conditions to ensure fair inclusion in recognition.''} (P6) \\
& \textit{``We had to introduce people of different skin tones and watched how the algorithm responded to each of them.''} (P7) \\
& \textit{``We seek people from different ethnicities, races, and genders to build a more diverse dataset.''} (P14) \\
& \textit{``We tried to set up a database that reflects Brazil's ethnic diversity to improve recognition.''} (P13) \\ \hline

\textbf{\textit{Testing for Model Consistency}} & \textit{``We started with exploratory tests initially, experimenting and seeing the answers that generated it.''} (P4)\\
& \textit{``We need to assemble this database so that the algorithm learns from those data, as bias can occur when data does not reflect reality.''} (P9) \\
& \textit{``We tried to find out if the AI's responses were consistent, or if it had variables that weren't in the original data.''} (P16) \\
& \textit{``We tested the modeling capacity of the model, verifying if it can present good results with data never seen.''} (P17) \\ \hline

\textbf{\textit{Testing for Data Representation and Balancing}} & \textit{``We greatly tested data balancing and how it influenced the results of the model.''} (P15) \\
& \textit{``We tried to ensure that the data we use to train are representative and balanced, seeking to avoid embedded bias.''} (P19) \\
& \textit{``We tested data balancing to see if the classes are well distributed, this is critical to avoiding bias.''} (P21) \\
\bottomrule
\end{tabularx}
\end{table}

Based on our analysis, we identified three distinct targets for fairness testing that practitioners actively use to detect biases and mitigate fairness issues. The first target, \textit{testing for diversity and inclusivity}, focuses on ensuring software performs fairly across diverse demographics, physical attributes, and cultural contexts. Practitioners discussing this target conduct detailed testing that includes sensitive attributes such as ethnicity, gender, and socio-economic background, validating that AI systems treat all demographic groups equitably. This user-centered approach highlights the social implications of fairness testing, aligning technical objectives with broader ethical considerations.

The second focus, \textit{testing for model consistency}, involves verifying the AI model's robustness, particularly when handling unfamiliar or unexpected inputs. Looking at this target, practitioners aim to ensure the model generalizes beyond its training data, producing stable and unbiased outputs across a range of conditions. This approach addresses the need for AI systems to function reliably in real-world scenarios, reducing the risk of inconsistent or biased responses. Professionals engaged in this testing strategy experiment with varied inputs, including edge cases, to validate model behavior across diverse situations.

Finally, \textit{testing for data representation and balancing} centers on ensuring that training data is representative and equitably distributed across user groups. Practitioners focusing on this target work to prevent bias at the data level by identifying overrepresentation or underrepresentation in key demographic categories. This approach is necessary for mitigating embedded biases that might otherwise go undetected, helping ensure that models do not disproportionately favor or disadvantage any particular group.

Among these three fairness testing strategies, \textit{testing for diversity and inclusivity} stands out as the most socially driven, prioritizing user-centered fairness. In contrast, \textit{testing for model consistency} and \textit{testing for data representation and balancing} are more technically oriented, focusing on algorithmic reliability and statistical fairness. Table \ref{tab:fairness_testing_targets} presents participant quotations illustrating these findings.

Beyond these specific testing targets, we identified a notable gap: \textit{testing for model behavior under extreme data changes}. This gap reflects the lack of systematic testing for how models respond to significant variations in data distribution, such as concept drift or drastic environmental changes, which can lead to inconsistencies and reduced reliability. Several participants acknowledged this limitation, noting that they had not thoroughly assessed the model’s response to extreme shifts in input data. 

For example, one participant stated, \textit{“We did not thoroughly test how the model would react to extreme changes in data distribution, such as large lighting variations”} (P17), while another noted, \textit{“We don’t examine the impact of extreme changes in data, such as concept drift, in all cases”} (P18). Others raised similar concerns, indicating that they did not investigate the effects of significant environmental variations, which may have impacted testing results (P15, P19). These findings highlight an area in need of attention to ensure fairness testing accounts for real-world variability.

\MyBox{\textbf{\emph{Finding 1 -- Summary.}} \small 
We identified three primary targets for fairness testing in AI systems: \textit{testing for diversity and inclusivity}, \textit{testing for model consistency}, and \textit{testing for data representation and balancing}. The first target ensures AI systems treat all demographic groups equitably by considering factors such as ethnicity, gender, and socio-economic background. The second verifies model robustness, ensuring consistent and unbiased performance across different inputs. The third focuses on preventing bias at the data level by ensuring diverse and balanced training datasets.}

\subsection{Fairness Testing in Practice: Strategies and Industry Approaches}

Based on our analysis, we identified emerging strategies for fairness testing in the industry, most of which are developed on an ad-hoc basis and lack formalized guidelines. Table \ref{tab:fairness_testing_strategies} summarizes these findings on how fairness testing is currently being implemented in the industry.

The first approach, \textit{continuous iteration and adjustment}, involves repeated testing and real-time model refinements. Practitioners cycle through test scenarios, gather feedback, and modify the model based on observed outcomes. Notably, this process is primarily led by data scientists, while testing professionals typically assist by defining scenarios rather than conducting the iterations themselves. 

\begin{table}[h!]
\caption{Strategies for Fairness Testing in Practice}
\centering
\label{tab:fairness_testing_strategies}
\begin{tabularx}{\linewidth}{p{2cm} X}
\toprule
\textbf{Strategy} & \textbf{Evidence Examples} \\ \hline

\textbf{\textit{Continuous \newline Iteration and \newline Adjustment}} & \textit{``The team did continuous iteration tests and analyzed the results to adjust the model in real-time.''} (P02) \\
& \textit{``We did continuous testing with user feedback, which helped refine the models.''} (P06) \\
& \textit{``To ensure the system worked correctly, we use many test rounds and adjustments based on users' feedback.''} (P05) \\
& \textit{``To validate the model, we made constant adjustments to the database and made several iterations with the customer.''} (P15) \\ \hline

\textbf{\textit{Manipulation and Simulation of Data}} & \textit{``To mitigate bias, we manipulated the images, changing shirts, positions, and even characteristics of the participants' hands.''} (P07) \\
& \textit{``When we couldn't manually augment the database, we created synthetic data, changing lighting and noise to extend the training set.''} (P09) \\
& \textit{``We also used simulated data in some parts to complement the samples where the actual data was not enough.''} (P15) \\
& \textit{``We use data increasing strategies such as GANs to create examples of people with characteristics that were not on the original dataset.''} (P18) \\ \hline

\textbf{\textit{Use of Well-defined Metrics}} & \textit{``We use metrics such as accuracy and precision to verify model performance.''} (P01) \\
& \textit{``We use A/B tests to verify that the model is presenting a bias for a particular group of users.''} (P17) \\
& \textit{``Another point is to use the right metrics. If the database is unbalanced, accuracy can mask the actual performance of the algorithm.''} (P18) \\
& \textit{``We tested models with metrics such as accuracy, recall, and precision to ensure that there are no apparent biases.''} (P19) \\ \hline

\textbf{\textit{Use of Specific Tools}} & \textit{``We used GANs data increase to generate examples of data that were not represented on the original dataset.''} (P19) \\
& \textit{``We made GPT adjustments so that it could better capture data diversity and avoid exclusion patterns.''} (P12) \\
& \textit{``We use a specific GPT to analyze patterns and similarities between approved projects, with the objective of facilitating new registrations.''} (P12) \\
& \textit{``We use data detox techniques to reduce bias, especially in negative connotations data.''} (P08) \\

\bottomrule
\end{tabularx}
\end{table}

The second approach, \textit{manipulation and simulation of data}, broadens test inputs to reflect real-world variations and capture edge cases. Manipulation involves altering data characteristics, such as changing user attributes or environmental conditions, to assess their impact on model performance. Simulation, on the other hand, generates synthetic data to represent conditions not present in the original dataset. Both techniques help enrich test coverage and completeness, often guided by testing professionals’ expertise.

The third approach utilizes \textit{well-defined metrics} to systematically evaluate fairness and identify bias. Data scientists and programmers apply metrics such as accuracy, recall, and precision to assess model outputs and ensure alignment with fairness objectives. Additionally, A/B testing is conducted to compare model behavior under varying conditions, enabling testers to identify biases affecting specific user groups within controlled experimental settings.

Lastly, practitioners leverage \textit{specialized tools} such as Generative Adversarial Networks (GANs) and GPT-based models to enhance data diversity, detect exclusion patterns, and strengthen model robustness. GANs are widely used for data augmentation, generating synthetic examples of underrepresented groups to improve fairness testing. GPT-based models, in turn, help analyze patterns and similarities in datasets, identifying trends that may contribute to model bias. These tools expand fairness testing by enabling practitioners to manipulate and simulate data more effectively, creating diverse and comprehensive testing scenarios.

Despite these emerging strategies, fairness testing in practice remains heavily reliant on verification processes conducted by data scientists and programmers, with limited involvement from dedicated testing professionals. This limited involvement occurs primarily because existing methods focus mainly on dataset design and model evaluation. As a result, critical decisions about fairness testing are often made without fully integrating specialized testing insights, leading to a narrower perspective on fairness-related issues. Additionally, industry-driven practices and tools specifically designed for fairness testing are still lacking, leaving many testing efforts inconsistent and ad-hoc.


\MyBox{\textbf{\emph{Findings 2 -- Summary.}} \small 
Fairness testing in industry relies on four key strategies, mostly developed on an ad-hoc basis due to the lack of formalized guidelines. Continuous iteration and adjustment involves repeated testing and real-time refinements, primarily by data scientists. Manipulation and simulation of data broaden test inputs by altering user attributes or generating synthetic data. Well-defined metrics, such as accuracy, recall, and A/B testing, help systematically detect biases. Specialized tools, including GANs and GPT-based models, enhance data diversity and reveal exclusion patterns. Despite these efforts, fairness testing remains largely dependent on data scientists and programmers, with minimal involvement from dedicated testing professionals.}

\subsection{Fairness Testing in Practice: Current Challenges}

We analyzed participants' narratives and identified recurring \textit{challenges in fairness testing} within the software industry (Table \ref{tab:fairness_testing_challenges} summarizes these findings). One of the most frequently reported obstacles was \textit{inadequate data quality and lack of diversity} in datasets. Many practitioners faced challenges when testing with unbalanced data and insufficient representation of specific demographic groups, limiting their ability to assess fairness effectively. Without diverse and representative datasets, testers struggle to validate models across varied user groups, increasing the risk of biases and fairness bugs in outputs. Thus, \textit{data quality and variability} remain pressing concerns in fairness testing.

Another major barrier is \textit{time constraints}, which restrict the scope of fairness testing and prevent comprehensive evaluations. The rapid pace of AI development and industry demands for swift feature releases often pressure teams to prioritize functionality over fairness testing. This time pressure inhibits in-depth analysis, allowing biases to go undetected and reducing the effectiveness of fairness assessments.

The \textit{absence of specialized fairness testing tools and metrics} further complicates the process. Practitioners emphasized that existing tools lack the ability to assess fairness beyond technical performance, making it difficult to evaluate societal implications. Without automated tools or well-defined metrics for bias detection, fairness testing remains labor-intensive, inconsistent, and difficult to scale across projects.

Challenges also stem from \textit{team composition and fairness knowledge gaps}. Many teams lack both diversity and specialized expertise in fairness testing, limiting their ability to identify and mitigate biases, particularly those tied to cultural, social, or demographic factors. Testing professionals often receive little or no formal training in fairness-related issues, making it difficult to meet the ethical and technical demands of AI projects.

Finally, \textit{model interoperability and black-box behavior} pose significant challenges. Practitioners struggle with AI models that generate \textit{unexpected or hallucinated outputs}, which can undermine fairness assessments. The lack of transparency in decision-making processes prevents testers from fully understanding or controlling model behavior, making it difficult to ensure fairness and explainability in AI systems.

\begin{table}[h!]
\caption{Challenges in Fairness Testing for Software Industry}
\centering
\label{tab:fairness_testing_challenges}
\begin{tabularx}{\linewidth}{p{2cm} X}
\toprule
\textbf{Challenge} & \textbf{Evidence Examples} \\ \hline

\textbf{\textit{Data Quality and \newline Diversity}} & \textit{``One of the major challenges was the limitation of time and data, which made it difficult to cover all test scenarios.''} (P03) \\
& \textit{``A big challenge was to get enough data to train AI and prevent it from learning based on irrelevant characteristics such as skin tone.''} (P07) \\
& \textit{``The lack of data on underrepresented groups caused challenges when testing the efficiency of the model for these groups.''} (P09) \\ \hline

\textbf{\textit{Time Constraints}} & \textit{``One of the biggest challenges was the time limitation to perform complete testing in all scenario variations.''} (P04) \\
& \textit{``There was a great pressure to deliver quick results, which impaired the critical analysis of the bias.''} (P12) \\
& \textit{``The pressure for rapid results prevented us from investigating bias issues in some situations in depth.''} (P18) \\ \hline

\textbf{\textit{Absence of \newline Specialized Tools and Metrics}} & \textit{``We didn’t always have clear metrics to measure performance, which made it difficult to identify bias issues.''} (P04) \\
& \textit{``We had problems with unlawful metrics, which generated a false perception that the model was performing well.''} (P14) \\
& \textit{``The lack of specific tools for testing AI completely was a challenge that limited the validation of the models.''} (P16) \\ \hline

\textbf{\textit{Team \newline Composition and \newline Knowledge Gaps}} & \textit{``The lack of diversity in the team has made it difficult to identify some more types of bias.''} (P02) \\
& \textit{``The composition of the team was a great challenge, as we did not always have people with the appropriate experience to identify certain biases.''} (P09) \\
& \textit{``The main challenge was dealing with the novelty of testing AI, something that many people on the team had no previous experience with.''} (P10) \\ \hline

\textbf{\textit{Model \newline Interoperability and Black-Box Limitations}} & \textit{``The AI hallucinated because it depends on the premises that have been programmed in the model, and this can be a problem when in production.''} (P04) \\
& \textit{``We had cases where it was hallucinating, creating completely meaningless answers, but at other times it seemed very reliable.''} (P16) \\
& \textit{``We already had problems with the hallucinating model, creating completely unrealistic answers in unexpected scenarios.''} (P17) \\

\bottomrule
\end{tabularx}
\end{table}

\MyBox{\textbf{\emph{Findings 3 -- Summary.}} \small 
We identified five major challenges in fairness testing: (1) Inadequate data quality and lack of diversity, where unbalanced datasets hinder bias detection; (2) Time constraints, limiting comprehensive fairness evaluations due to industry pressure for rapid releases; (3) Absence of specialized tools and metrics, making fairness testing labor-intensive and inconsistent; (4) Team composition and fairness knowledge gaps, as many teams lack the expertise to identify and mitigate biases; and (5) Model interoperability and black-box behavior, complicating transparency and fairness assessments.}

\section{Discussions}  \label{sec:discussions}
We analyzed our findings alongside existing literature to assess the current state of software fairness in both research and industry practices, as well as to explore the implications of these insights.

\subsection{Software Fairness Testing: The Gap Between Academic Research and Practice}

While researchers have made significant contributions in defining individual and group fairness by introducing concepts such as fairness through unawareness, awareness, counterfactual fairness, and causal fairness for individual fairness, and statistical parity, equalized odds, and equal opportunity for group fairness \cite{chen2024fairness, soremekun2022software, zhang2021ignorance}, these advancements are yet to be fully integrated into industry practice. In theory, these definitions offer structured approaches for ensuring equitable outcomes across different demographic groups and for making fair individual comparisons, which could ideally serve as the foundation for standardized testing processes. However, our findings reveal that, in practice, professionals still rely on ad-hoc methods to implement fairness testing due to a lack of clear, actionable guidelines. This gap between theoretical frameworks and real-world applications highlights the urgent need for practical, standardized tools and guidelines to operationalize fairness definitions within the software industry.

The literature offers a structured, component-based approach to fairness testing in machine learning software, covering aspects such as data, algorithm implementation, processes and configurations, model behavior, and non-ML components (e.g., data storage and user interfaces that influence fairness), with methods designed to detect fairness issues at each stage \cite{chen2024fairness}. However, our findings indicate a significant divergence in industry practices, where fairness testing often relies on adapting basic testing methods to fit specific project needs without clear distinctions between the levels or types of components. In practice, participants typically focus on data and model behavior, generally limited to immediate, observable outcomes, with little attention to systematic testing approaches across ML components.

The literature shows a growing number of open-source fairness testing tools available for ML applications, with several tools identified across domains like general ML, deep learning, natural language processing, computer vision, and speech recognition \cite{chen2024fairness}. However, limitations such as steep learning curves, insufficient documentation, limited configurability, and lack of regular maintenance hinder their adoption in real-world industry settings \cite{nguyen2024literature}. Our findings echo these issues, revealing that, despite the documented tools in literature, these resources are rarely incorporated into everyday fairness testing practices in the software industry. This gap between tool availability and practical usage demonstrated the need for tools that are more aligned with industry demands and practitioner needs. In particular, given the knowledge gaps around fairness testing methods reported by practitioners, tools designed for ease of use could play a significant role in making fairness testing more accessible and actionable within industry practices.

\vspace{-0.5em}
\subsection{Implications for Research}
While our findings confirm the persistent gap between fairness theory and its practical implementation, they also point to several research directions to bridge this divide. Theoretical progress in defining fairness has introduced structured approaches such as fairness through awareness, counterfactual fairness, and statistical parity. However, these definitions remain challenging for practitioners to apply due to their complexity and lack of direct integration into existing software development workflows. To address this, research should prioritize the development of frameworks that translate fairness definitions into actionable methodologies that software teams can incorporate seamlessly into testing processes.

Moreover, our study highlights the need for comprehensive, industry-aligned fairness testing tools that go beyond academic prototypes. While the literature has produced numerous tools for bias detection and fairness validation, their adoption in industry remains limited due to high technical barriers, lack of configurability, and insufficient documentation. Research should focus on designing user-friendly, well-documented, and scalable tools that integrate with industry-standard workflows. Additionally, empirical studies assessing the effectiveness of fairness testing methodologies in real-world settings could provide valuable results into best practices and gaps that require further attention.

Another critical area for research is the automation of fairness testing processes. Given the time constraints and rapid development cycles in industry, manual fairness assessments are often deprioritized. Investigating how fairness testing can be integrated into continuous integration and deployment pipelines could lead to more scalable solutions. Future research should also investigate how fairness testing interacts with broader software quality assurance practices, ensuring that fairness considerations become an integral part of software development rather than an isolated concern.

\subsection{Implications for Practice}
From an industry perspective, our findings suggest that fairness testing will likely remain fragmented and inconsistently applied unless concrete steps are taken to embed it into routine software development practices. Currently, fairness assessments rely heavily on ad-hoc methods, with significant variations across projects. To move beyond this fragmented approach, organizations should invest in standardized fairness testing practices that align with existing development and testing workflows. This requires clear, actionable guidelines that help practitioners incorporate fairness testing without adding significant overhead.

Training and education also emerge as critical factors in improving fairness testing adoption. Many software teams lack the necessary expertise to identify and mitigate biases, particularly in contexts where fairness concerns extend beyond technical aspects to social and ethical considerations. Providing targeted training for software developers, testers, and data scientists could help bridge this knowledge gap. Furthermore, companies should consider establishing dedicated roles or cross-functional teams that combine expertise in software testing, AI ethics, and social impact analysis to ensure more comprehensive fairness assessments.

Another key consideration for industry is the integration of fairness testing tools into standard testing pipelines. As our findings indicate, fairness testing is currently performed predominantly by data scientists and programmers, with limited involvement from dedicated testing professionals. To ensure fairness testing becomes a sustainable practice, organizations should adopt tools that enable seamless integration into existing automated testing environments. Additionally, fostering collaboration between different roles (e.g., software testers, data scientists, and domain experts) can enhance the effectiveness of fairness assessments and lead to more robust, bias-aware AI systems.

Finally, as regulatory frameworks around AI fairness continue to evolve, industry practitioners should take a proactive approach in shaping fairness standards and compliance mechanisms. Rather than waiting for external enforcement, companies can lead the way by developing internal policies and industry-wide best practices. Establishing partnerships between academia and industry could further accelerate the development of practical solutions, ensuring that fairness considerations are embedded throughout the software lifecycle rather than addressed as an afterthought.

\subsection{Threats to Validity}
Our findings should be interpreted with awareness of the limitations of the qualitative methodology used in this study, which emphasizes criteria such as credibility, resonance, and practical relevance over quantitative measures such as statistical validity or generalizability. We ensured credibility by establishing a clear chain of evidence linking participants’ experiences to our conclusions, as shown in the tables \ref{tab:fairness_testing_targets}, \ref{tab:fairness_testing_strategies} and \ref{tab:fairness_testing_challenges}. Our sample included participants from diverse professional backgrounds, allowing for a range of perspectives, and resonance was achieved by involving participants in refining and validating findings throughout the analysis. The strength of our research lies in its focus on real-world challenges in fairness testing and the practical gaps in current tools and practices.

As is typical in qualitative research, our findings are not intended for statistical generalization across all software industry contexts. Instead, our in-depth interviews reached data saturation, providing detailed insights into participants' experiences with fairness testing. These insights are best seen as transferable to similar settings rather than universally generalizable. We have provided detailed descriptions of the study context and participants to support re-analysis and applicability in comparable contexts. In the end, while we identify correlations, such as the influence of tool availability on fairness testing practices, these findings should be cautiously applied to broader industry practices.

Still, regarding the qualitative nature of our study, another limitation worth noting was the use of observational note-taking during interviews to capture behaviors such as hesitation, emphasis, or emotional cues. While these memos were useful for contextualizing participant responses, they introduce a degree of subjectivity and limited replicability. Different researchers may interpret the same non-verbal cues in inconsistent ways, which poses a potential threat to validity. These observational notes were not included in the formal coding process but served as contextual supplements during interpretation, helping the research team identify moments of ambiguity or emphasis that may inform thematic insights.

Finally, in this study, we used GPT-4 Omni through prompt engineering to assist with open-coding, yet this approach introduces some limitations. The effectiveness of prompt engineering depends on clarity and precision in crafting prompts, which may risk introducing bias in the qualitative analysis process. Although the few-shot examples and a defined task persona helped direct the model, the inherent variability of generative AI models presents challenges for the consistency and reproducibility of results. We mitigated this issue by conducting a manual accuracy check to verify alignment between the codes generated by the model and the original transcripts. 
This approach allowed us to maintain consistency and accuracy in our findings while leveraging the model's efficiency for initial coding.

\section{Conclusions}  \label{sec:conclusions}
Ensuring fairness in AI-powered systems has become a growing concern within the software industry, particularly as biases in machine learning models can lead to discriminatory outcomes. Despite significant advancements in fairness research, there is limited understanding of how fairness testing is applied in real-world development settings. This study aimed to bridge this gap by investigating how software professionals test AI systems for fairness, identifying the practices currently in use, the challenges faced, and the areas requiring further development. Through interviews with 22 industry professionals working on AI and machine learning projects, we explored the extent to which fairness testing is integrated into software development workflows and the factors influencing its implementation.

Our findings reveal three primary targets of fairness testing in industry: diversity and inclusivity, model consistency, and data representation. While diversity and inclusivity testing ensures that AI models treat all demographic groups fairly, model consistency testing focuses on verifying robustness across different inputs, and data representation testing aims to mitigate biases stemming from unbalanced training datasets. In terms of how fairness testing is performed, practitioners primarily rely on iterative adjustments, data manipulation, and metrics-based evaluations, often using specialized tools to enhance test coverage. However, we identified key challenges, including poor data quality and representation, time constraints, lack of dedicated fairness testing tools, and gaps in practitioner expertise, which limit the effectiveness of fairness testing in practice. These challenges highlight the urgent need for structured methodologies and industry-aligned fairness assessment frameworks.

The implications of our study extend to both research and practice. Academically, our findings emphasize the gap between fairness theory and its implementation, calling for the development of practical, industry-adopted frameworks and tools that can facilitate fairness testing in real-world software projects. For industry, the lack of standardized fairness testing procedures and automated solutions suggests the need for greater collaboration between researchers and practitioners. Establishing cross-functional teams that integrate ethical considerations into software testing workflows could also help improve fairness outcomes and bridge the knowledge gaps observed in industry teams.

As immediate future work, we aim to develop strategies to enhance fairness test case creation and refine fairness testing oracles to better support practitioners in identifying and mitigating biases. This will involve designing more systematic approaches for fairness validation, ensuring that fairness testing evolves from an ad-hoc activity to an essential component of software quality assurance. Additionally, in our future research we plan to explore ways to integrate fairness testing more seamlessly into existing CI/CD pipelines, fostering more consistent and scalable fairness assessments in AI-driven software development.

\section{Data Availability}
Anonymized interview quotations are available at \url{https://figshare.com/s/d6676bece29428ee7a25}. Quotes with identifying details were excluded for privacy, so some participants may not have associated quotations in the spreadsheet. Finally, some phrasing may reflect direct translations from other languages. 
 
\section*{Acknowledgments}
This work was supported by the Natural Sciences and Engineering Research Council of Canada (NSERC), Discovery Grant RGPIN-2024-06260, and by Alberta Innovates through the Advance Program, Project Number 24506125.

\ifCLASSOPTIONcaptionsoff
  \newpage
\fi

\balance
\bibliographystyle{IEEEtran}
\bibliography{Main.bib}

\end{document}